# INDENTATION, ELASTICITE ET TENSION DE SURFACE

## Christophe Fond
*Institut de Mécanique des Fluides et des Solides, Université de Strasbourg*

Les modèles classiques de Hertz, Sneddon et Boussinesq fournissent des solutions pour les problèmes de l'indentation d'un massif élastique semi-infini par respectivement une sphère, un cône et un poinçon plat. Bien que ces modèles aient été largement éprouvés, il s'avère qu'à petites échelles et pour des matériaux souples, la tension de surface peut contribuer considérablement à la réponse mécanique à l'indentation. Les échelles sont typiquement celles inférieures au micron pour un élastomère et celles inférieures au millimètre pour un gel. L'exploitation de certains résultats expérimentaux de microscopie ou de nanoindentation demeurent approximatifs en raison de l'absence de modèles intégrant l'effet de la tension de surface. Afin de n'avoir qu'un seul paramètre pour définir l'élasticité − le module de cisaillement $\mu$ - en regard de la tension de surface $\gamma$, le module de compressibilité est fixé à 2 GPa pour tous les calculs suivants. Ces résultats sont donc dédiés aux milieux quasi-incompressibles. La tension de surface peut être considérée constante pour des rayons de courbures supérieurs au nanomètre [Fisher et Israelachvili, 1980]. Pour les calculs suivants, comme schématisé en Fig. 1, elle peut être commodément considérée comme une précontrainte en surface à condition que les calculs intègrent les non linéarités géométriques des grands déplacements [Adamson, 1990].

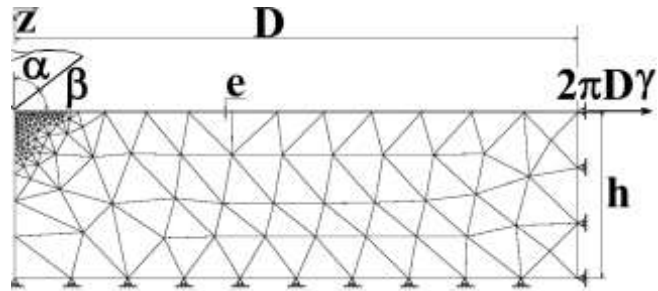

*Fig. 1. Vue schématique d'un maillage axisymétrique montrant l'indentation d'un substrat par un cône et une "peau de tambour" tendue par une précontrainte $\gamma$.*

La formulation par la méthode des éléments finis (MEF) intègre donc les grands déplacements. La somme des forces généralisées intérieures et extérieures peut être exprimées par [Zienkiewicz et Taylor, 1991] :

$$\Psi(u) = \int_V [B_0 + B_L(u)]^T \sigma \, dV - f = 0 \qquad (\text{éq. 1})$$

où u représente les déplacements nodaux, f les forces nodales élémentaires, $B_0$ la matrice de initiale de transformation des déplacements en déformations, i. e. $\partial\varepsilon = B \, \partial u$ où $\partial\varepsilon$ est l'incrément de déformations et $\partial u$ celui de déplacement, $B_L(u)$ le changement dans la matrice de transformation lié aux déplacements nodaux actuels. Considérant que $\partial f = 0$, l'équilibre se traduit par :

$$\partial\Psi(u) = \int_V [\partial B_L]^T \sigma + \int_V [B]^T \partial\sigma \, dV = 0 \qquad (\text{éq. 2})$$

où $\partial\sigma = C \, \partial\varepsilon$, C représentant le tenseur d'élasticité usuel et $B = B_0 + B_L$. La matrice $B_L$ peut être linéarisée et le premier terme de l'intégrale de l'éq. 2 devient dépendant de du. L'équation 2 peut donc être réécrite sous la forme :

$$\partial\Psi(u) = K_\sigma \, \partial u + [K_0 + K_L] \, \partial u = 0 \qquad \text{(éq. 3)}$$

où $K_\sigma \, \partial u \equiv \int_V [\partial B_L]^T \sigma$, $[K_0 + K_L] \, \partial u \equiv \int_V [B_0 + B_L]^T C [B_0 + B_L] \, \partial u \, dV$. Pour nos applications, le niveau des contraintes initiales dans les éléments de surface est plusieurs ordres de grandeur supérieur à celui induit par l'indentation dans le substrat. En effet, la contrainte totale est $\sigma = C(\varepsilon - \varepsilon_0) + \sigma_\gamma$, où $\varepsilon_0 = 0$ en élasticité linéaire et $\sigma_\gamma \gg C\varepsilon$. En pratique, la rigidité tangente totale est donc quasiment égale à la rigidité initiale à laquelle s'ajoute la matrice de non linéarité géométrique $K_\sigma$.

$$\partial\Psi(u) = [K_\sigma + K_0 + K_L] \, \partial u \approx [K_\sigma + K_0] \, \partial u \qquad \text{(éq. 4)}$$

Il faut préciser que, mise en oeuvre de façon classique dans le logiciel par éléments finis [CAST3M$^\copyright$], la précontrainte $\sigma_\gamma$ est liée au repère $(r, z)$ initial et non au repère actuel de l'élément de sorte que la composante de tension alignée avec l'axe r est – pour autant que les incréments soient effectivement infinitésimaux – constante quelle que soit la rotation de l'élément de surface. Un schéma d'intégration implicite dit "$\theta$-method" est utilisé pour calculer la solution. Entre chaque pas de calcul, la précontrainte est actualisée pour rester constante parallèlement à la surface.

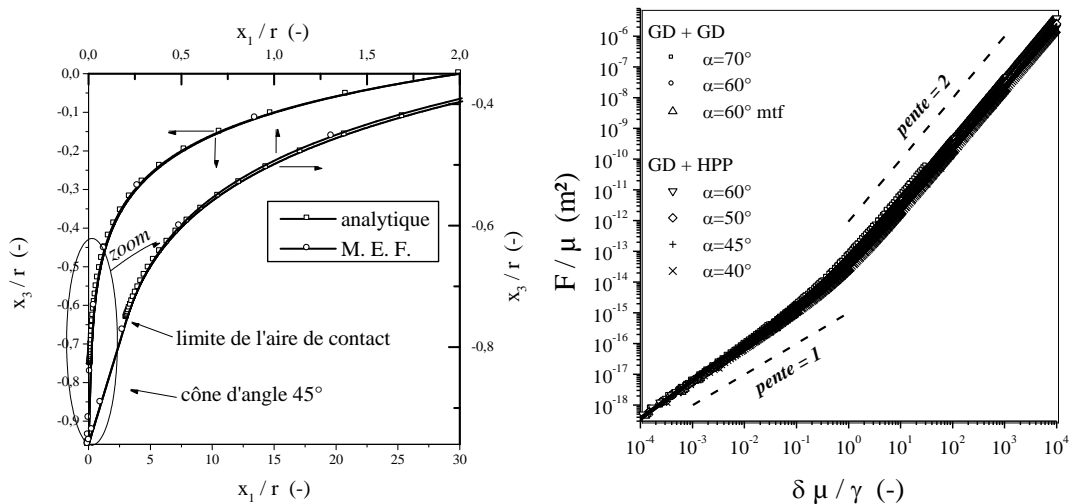

*Fig. 2. (gauche) comparaison des déformées de la surface prédites par la solution analytique et par la MEF. (droite) Courbes forces vs. profondeur d'indentation pour des indenteurs coniques dont $\alpha = 40$ à $70°$.*

La Fig. 2 de gauche montre que les résultats obtenus par la MEF sont convenables jusque dans le cas extrême où la réponse volumique s'annule et seule reste la réponse surfacique. La Fig. 2 de droite montre l'effet de la tension de surface aux petites échelles, typiquement pour $\delta\mu/\gamma < 1$. Comme attendu, la force varie avec le carré de l'enfoncement pour la solution de Sneddon. En revanche, comme prédit par la solution analytique valant pour seule une membrane tendue, la force tend progressivement à varier linéairement avec l'enfoncement aux petites échelles.

Des résultats analogues ont été obtenus pour les solutions de Hertz et Boussinesq. Il est ainsi possible de montrer que l'on peut - en pratique - toujours superposer la solution de Boussinesq à celle de Hertz ou Sneddon pour simuler une adhésion lors d'un retrait d'indenteur. Ainsi, tous ces résultats permettent de réévaluer les aires spontanées de contact prédites par la théorie de JKR. En présence de tension de surface.

**Références**